# Photoluminescence Mapping over Laser Pulse Fluence and Repetition Rate as a Fingerprint of Charge and Defect Dynamics in Perovskites

Shraddha M Rao,[1] Alexander Kiligaridis,[1] Aymen Yangui,[1] Qingzhi An,[2] Yana Vaynzof[2,3] and Ivan G. Scheblykin[1*]

[1]*Chemical Physics and NanoLund, Lund University, P.O. Box 124, 22100 Lund, Sweden*

[2] *Chair for Emerging Electronic Technologies, Technical University of Dresden, Nöthnitzer Str. 61, 01187 Dresden, Germany*

[3] *Leibniz-Institute for Solid State and Materials Research Dresden, Helmholtzstraße 20, 01069 Dresden, Germany*

*)**Corresponding Author:** *ivan.scheblykin@chemphys.lu.se*

Defects in metal halide perovskites (MHP) are photosensitive, making the observer effect unavoidable when laser spectroscopy methods are applied. Photoluminescence (PL) bleaching and enhancement under light soaking and recovery in dark are examples of the transient phenomena that are consequent to the creation and healing of defects. Depending on the initial sample composition, environment, and other factors, the defect nature and evolution can strongly vary, making spectroscopic data analysis prone to misinterpretations. Herein, the use of an automatically acquired dependence of PL quantum yield (PLQY) on the laser pulse repetition rate and pulse fluence as a unique fingerprint of both charge carrier dynamics and defect evolution is demonstrated. A simple visual comparison of such fingerprints allows for assessment of similarities and differences between MHP samples. The study illustrates this by examining methylammonium lead triiodide (MAPbI$_3$) films with altered stoichiometry that just after preparation showed very pronounced defect dynamics at time scale from milliseconds to seconds, clearly distorting the PLQY fingerprint. Upon weeks of storage, the sample fingerprints evolve toward the standard stoichiometric MAPbI$_3$ in terms of both charge carrier dynamics and defect stability. Automatic PLQY mapping can be used as a universal method for assessment of perovskite sample quality.

## 1. Introduction

Metal halide perovskites (MHPs) are an important class of semiconductors, that received significant attention over the last decade due to their excellent optoelectronic properties and promising performance in various optoelectronic devices.[1-3] Methylammonium lead triiodide (MAPbI$_3$ or MAPI) has since proven to be an excellent representative of the versatility of perovskites and has emerged as a promising photovoltaic material. Understanding charge carrier dynamics and defect physics in these semiconductors is critically important for advancing the progress in making stable devices with reproducible properties for optoelectronic applications.[4-9] Despite extensive studies, many contradicting observations have been reported that continue to puzzle researchers. The main challenges of working with perovskites are: i) poor sample storage stability, ii) poor reproducibility in sample fabrication,[10, 11] and iii) evolution of the material under light irradiation, current flow, and electric field[12-18] and material self-healing.[19-22] Most of these difficulties, as well as advantages like, e.g., self-healing originate from low energy of defect formation, easiness of material recrystallization,[5] the ionic nature of the defect states and their metastability[22-26] due to very efficient ion migration in MHPs.[15, 27, 28]

Unreliable performance of MHPs correlates with the extreme sensitivity on the specific details of the fabrication process,[10, 29-32] one of them being stoichiometries of precursor solutions.[33, 34] Previous studies have shown how unintentional and fractional discrepancies as low as 0.5–1% in precursor ratios lead to samples with very different surface energetics and emission properties,[11] even when their absorbance and film morphology remain unchanged.

Photoluminescence (PL) spectroscopy is a very powerful and popular tool to rationalize MHPs due to extreme sensitivity of PL properties to defect states.[8] Recently a unique 2D photoluminescence quantum yield mapping technique (PLQY($f$,P) mapping, where $f$-pulse repetition rate and P-pulse fluence) to characterize charge carrier dynamics in polycrystalline MAPI films was demonstrated.[35] This new technique provides extensive information to unravel the various photophysical processes that occur with charge carriers in luminescence semiconductors like radiative recombination, trapping, trap-assisted non-radiative recombination, Auger recombination. The amount of data is sufficient for extracting the rate constants and trap densities by applying appropriate theoretical models to fit the PLQY map and PL decay kinetics. However, it was shown that even for the classical, rather stable MAPI films it is not possible to fully explain the observed PLQY($f$,P) maps and PL decay kinetics measured from the same sample using an extended Shockley-Read-Hall (SRH+) model of charge recombination, the reason for this is not yet known.

It was proposed that the PLQY(*f*,P) map can be used as the sample fingerprint[16, 35] because it reflects the peculiarities of charge carrier dynamics at very different excitation conditions in terms of charge carrier concentration created by one laser pulse (pulse fluence ranges over four orders of magnitude) and the time gap between laser pulses (ranges over six orders of magnitude from 10 ns to 10 ms).

In this work we concentrate on a very important aspect of MHPs that is, although known for practicing laser spectroscopists,[8] usually not in focus of publications discussing charge carrier dynamics extracted from laser spectroscopy experiments. We talk about defect state instability under light and defect state evolution over the sample age. This can lead to irreproducibility of experimental data and so-called "observer effect", which means that the experiment itself changes the sample. The latter also means that the apparent results of the experiment depend on the particular experimental details which at first glance should not matter, for example the order of measured data points, waiting time between consecutive measurements and other factors. The importance of the observer effect should not be overlooked because most probably it is present in all published data on MHPs to one extent or another and can lead to the proposition of various theories and explanations which in fact might be a consequence of measurement induced artefacts. Because the PLQY(*f*,P) mapping is fully automated, it is ideal for comparison of samples possessing the observer effect, because the experimental conditions can be kept absolutely identical for all samples.

We measured the PLQY(*f*,P) maps of MAPbI$_3$ films prepared with fractional variations in the precursor stoichiometry. We focus on exploring overstoichiometric films (iodide fraction higher than three) as those were shown to result in a superior photovoltaic performance.[11] The PLQY maps reveal sensitive changes in the sample PL and stability over time. We found the presence of reversible light-induced fast dynamics in the non-stoichiometric MAPI pointing to the dependence of the defect state concentration on the history of light exposure of the samples. These fast defect kinetics at times scales of seconds are easily hidden from conventional spectroscopic investigations. We also observe maturing/healing of the samples over their shelf storage time over weeks leading to a more stable PL response and PLQY maps close to that of the standard MAPI$_3$. The advantage of the PLQY mapping is such that all these features are easily detected by simple visual comparison of the PLQY maps making this technique highly useful for monitoring of the perovskite sample quality.

2. **Experimental Section**

The samples were synthesized using a previously reported method, where the precursor stoichiometry was carefully controlled by varying the ratio between MAI and Pb(OAc)$_2$ to yield MAPI films with fractionally varying stoichiometries: stoichiometrically balanced MAPbI$_3$, slightly over-stoichiometric MA$_{1.04}$PbI$_{3.04}$ and significantly over-stoichiometric MA$_{1.1}$PbI$_{3.1}$ (See Note S1, Supporting Information, for details). They were stored in a nitrogen environment, isolated from light exposure before and between measurements. The PL of the samples was excited and imaged using a home-built wide-field photoluminescence microscope based on Olympus IX71 and an EMCCD Camera (ProEM, Princeton Instruments). The samples were excited using a 485 nm diode laser (PicoQuant), controlled by a multichannel laser driver (SEPIA LD828, PicoQuant).

The PLQY mapping entails exciting the sample with a pulsed laser at repetition rate $f$ [s$^{-1}$] and pulse fluence P [photons/cm$^2$/pulse] that were controlled by the laser driver and a neutral density filter wheel, respectively. For each combination of $f$ and P and PL intensity (PL image for the setup) of the sample was measured. Because the entire system was calibrated, this data was later converted to the external PLQY (see Note S2, Supporting Information). To obtain the PLQY($f$,P) map, PL was measured for a laser excitation spanning five orders of magnitude (from *ca*. 10$^8$ to 10$^{12}$ photons/cm$^2$/pulse) in five steps with power fluences P1, P2, P3, P4, and P5 (each step changes the fluence approximately ten times) and almost seven orders of magnitude in pulse repetition rate, i.e., from 100 Hz to 80 MHz with the repetition rate changed by ≈2–3 times each step. For each pulse fluence P, the repetition rate, $f$, was scanned across the entire range. The final PLQY($f$,P) map was plotted as PLQY versus the time averaged power density, W, which is defined as $W$ = photon energy·$f$·P with units W cm$^{-2}$. In total the map consists of ≈80 data points, see Note S3 (Supporting Information) for details.

Data acquisition was designed so that all data points for the entire PLQY($f$,P) map were acquired automatically (**Figure** 1a). All components of the setup were controlled by a LabVIEW program that executes the experiment according to a pre-loaded table of parameters of the data acquisition ($f$, P, shutter timing, acquisition time of the camera, filters, etc.). To avoid unnecessary sample exposure, the shutter was synchronized with camera to allow the laser beam reaching the sample during the PL acquisition only. The program automatically saved the PL images, which were later processed using another program to yield the PLQY map. This ensured that the data acquisition conditions were absolutely reproducible meaning that the same experiment could be repeated with another sample. Automation also excludes human errors that are unavoidable when such long series of data at varied conditions are measured. The entire measurement took from 1 to 3 h, where the longest time took to acquire data for low P and low $f$ values since exposure times as long as several minutes per data point were often required. In our particular case the total exposure of the sample was ≈19 J cm$^{-2}$ that, however, was spread over ≈1 h of the experiment.

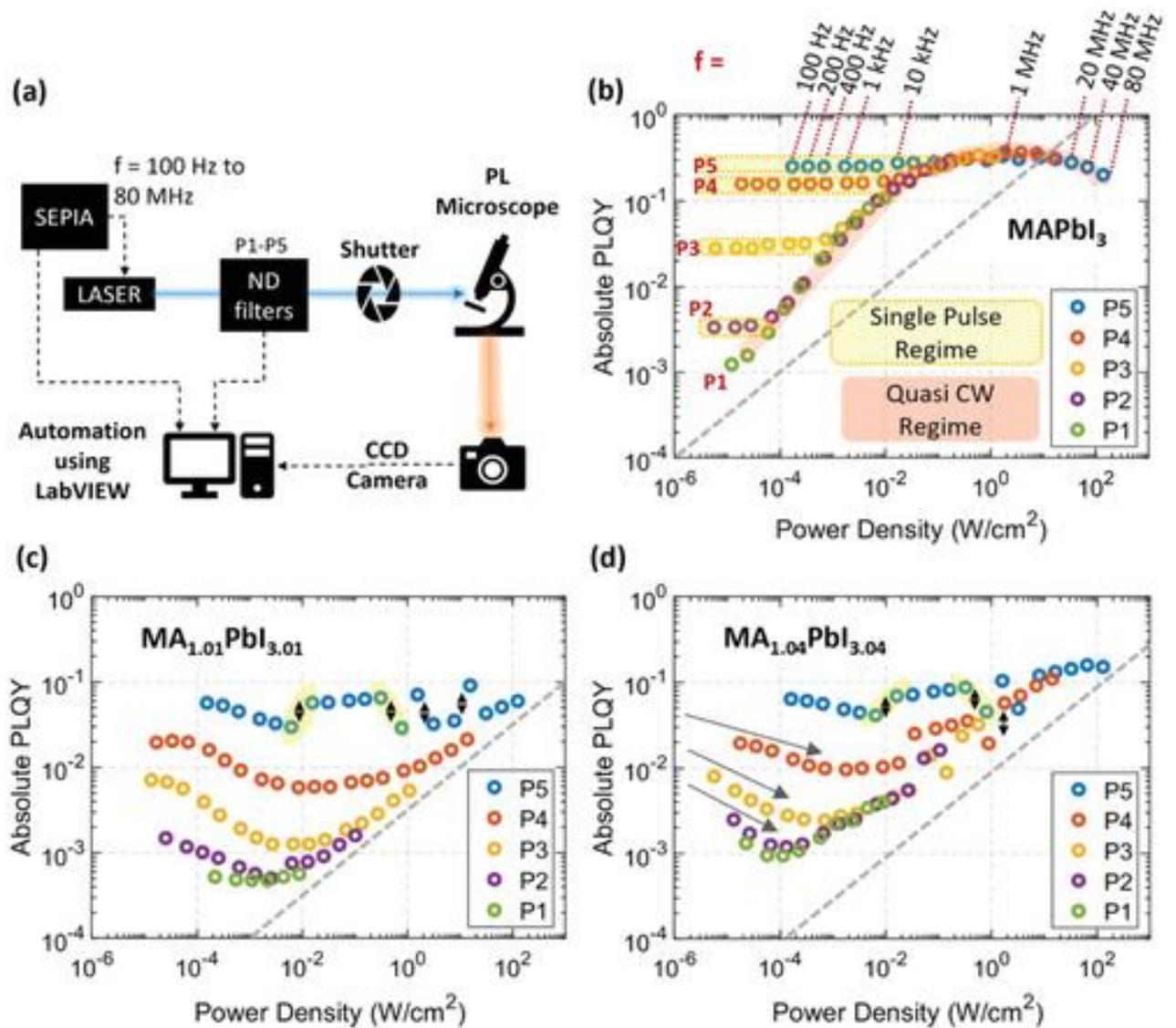

**Figure 1.** a) A schematic representation of the setup for acquisition of PLQY(*f*,P) maps. PLQY maps of b) MAPbI$_3$, c) MA$_{1.04}$PbI$_{3.04}$, and d) MA$_{1.1}$PbI$_{3.1}$ films. The map is formed by plotting together a family of 5 curves of PLQY versus Power Density. Each curve is obtained for its own pulse fluence, Pn. P1 is the lowest and P5 is the highest pulse fluence (in photons/cm²/pulse). The scanning of the laser repetition rate *f* is illustrated in (b) for pulse fluence P5. See the text for details. The grey dash lines show the square root dependence between the excitation power density and PLQY, which aids to compare the slopes of the quasi-CW-regime of the samples. Different excitation regimes are marked in (b).

3. **Results and Discussion**

*3.1 PLQY Map of the Standard MAPbI$_3$ and Explanation of the Excitation Regimes*

First, let us consider the PLQY(*f*,P) maps of the fresh samples, as shown in Figure 1. A family of PLQY versus excitation power density curves, one for each pulse fluence (P1, P2, etc) are plotted together to give the final PLQY(*f*,P) map. The map of the stoichiometric MAPI film is quite close to the one previously reported.[35] The over stoichiometric samples show very differently looking maps: PLQY is lower at all conditions and the shape of the features of the map is unusual. Already here the visual comparison of the PLQY maps reveals that these samples are very different in terms of their PL and response to light irradiation in comparison to the stoichiometric MAPI.

The PLQY map of the standard MAPI is used here to recall the signatures of the different excitation regimes and charge recombination channels in semiconductors.[35] The data points on the PLQY map can be distinguished as belonging to two excitation regimes: i) the single pulse regime that is marked in yellow on Figure 1a, where the PLQY data points for fixed P follow a straight horizontal line when f is changed; ii) the Quasi-CW regime that is marked in orange on Figure 1a, where we observe merging of the individual lines measured at fixed P into one common dependence where PLQY depends the averaged power density $W$ = photon energy·*f*·P only. In the single pulse regime, the PL excited by one laser pulse is free from the memory of the excited state dynamics produced by the previous pulse. In the quasi-CW regime, to the contrary, the PL excited by the first pulse affects the evolution of the PL induced by the next pulse because the time interval between two consecutive pulses is too short to ensure complete relaxation of all transient species (free and trapped charge carriers) created by the previous pulse. The existence of these two regimes is rather obvious and discussed in detail elsewhere.[35]

The PLQY map of our reference sample, plotted in the manner as shown in Figure 1a, with some imagination resembles a "horse neck with mane", where the "neck" corresponds to the quasi-CW regime, and the "mane" is formed by the separate strands of single pulse regime data points collected for fixed values of P (P1, P2,…, P5).[35] These features of the PLQY maps will be referred to in the following sections of the article to explain, analyze and compare the MAPI film samples.

The dotted grey line shows a square root dependence of PLQY on the average excitation power density (slope 0.5 in the log-log plot) expected for the quasi-CW regime in the case of photodoping (SRH model without trap saturation). The line is used as a reference for the slopes of the quasi-CW regime. The map of the stoichiometric MAPI film shows a quasi-CW slope of ≈0.77 that was previously explained by a partial trap saturation,[35] and the single pulse regime forming well-separated, almost horizontal lines for different pulse fluences. This PLQY map can be reasonably well explained by an extended SRH charge recombination model.[35]

*3.2 PLQY Maps of the Overstoichiometric MAPI and Evidence of the Observer Effect*

As seen in Figure 1c,d, the PLQY maps of overstoichiometric MAPI samples differ substantially from the standard map. First, the maps are shifted down in comparison with the reference, indicating a lower absolute PLQY at all excitation regimes. Second, the slope of the quasi-CW regime is reduced to 0.5 for the slightly over-stoichiometric MAPI (Figure 1c) and to <0.5 for the significantly over-stoichiometric MAPI sample (Figure 1d). Third, for both these samples the single-pulse regimes show an uncharacteristic initial reduction of PLQY as *f* increases (marked by three grey arrows in Figure 1d) and then a recovery to join the quasi-CW "horse neck". Changes of PLQY in the single pulse regime are not consistent with any charge recombination model where the defect concentration and trapping rates are static. Note that these maps were highly reproducible when measured multiple times for each different sample, at nine different locations within the samples. So, we must investigate the possibility that the data acquisition itself changes the charge recombination dynamics, or in order words, we need to examine the presence of the observer effect.

Another highly unusual feature clearly visible in the PLQY maps of $MA_{1.04}PbI_{3.04}$ and somewhat less pronounced for $MA_{1.1}PbI_{3.1}$ sample, is an apparent "discontinuity" in the single-pulse regime ("horse mane") for the highest pulse fluence P5, as marked by yellow ovals highlighted with double-sided arrows. This feature was reproduceable, irrespectively in the particular sample or the excitation spatial location. Even multiple measurements at the same region yielded the same peculiar jumps in the PLQY values by a factor of two without any obvious reason. As we will show below, this feature has the same origin as the previously discussed "curved single pulse regime": it is the observer effect, related to photosensitivity of the material and the exact details of sample light exposure controlled by a shutter and the time window over which the PL photons are acquired by the camera at the sub-second timescale.

*3.3 PL Response on Switching the Laser, Reversible Bleaching*

Before we explain the reasons for these unexpected observations, let us discuss the time-resolved response of the PL intensity on switching the laser excitation on (**Figure** 2). During this experiment the PL signal from $MA_{1.04}PbI_{3.04}$ sample was recorded constantly with 0.03 s time resolution while the shutter blocking the laser beam was opened and closed multiple times. The PL intensity is high at the start of each excitation period, but quickly declines by approximately a factor of two, reaching a steady state value. Remarkably, after a few seconds in dark, the sample shows exactly the same PL response when excited again, meaning that the observed PL bleaching (or decline) under illumination is fully reversible. This reversible PL dynamics at timescale from milliseconds to tens of seconds (depending on the excitation power density) must affect the PL intensity measurements for PLQY mapping.

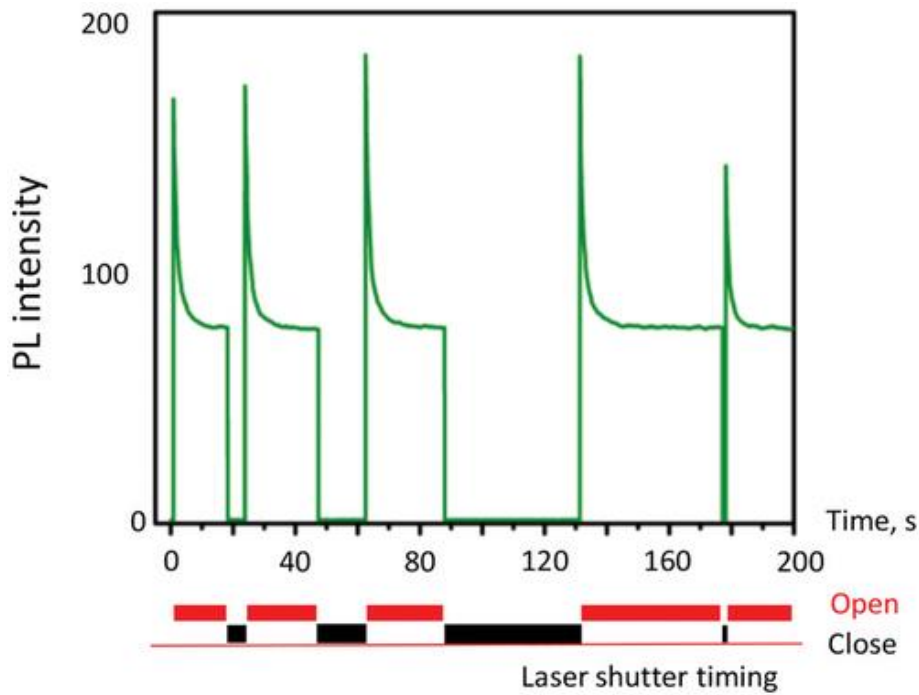

**Figure 2.** Evolution of PL of MA1.04PbI3.04 sample under interrupted laser irradiation (pulse repetition rate 10 MHz, pulse fluence P5). PL is bleaching at the time scale of seconds, however, after a short rest in dark, the PL intensity fully recovers.

An important point to note is that for measuring of a PLQY map the data acquisition is programmed for a minimum, but sufficient acquisition time. For a low power density excitation, the PLQY is very low and the acquisition times as long as 10–300 s are needed, whereas for high excitation power densities, acquisition times from 0.07 to 1 s are sufficient. Considering the fast bleaching noted above, we expect the PL intensity to change during the acquisition time of the data points of the PLQY($f$,P) map.

*3.4 Quantitative Assessment of the Effect of Temporal PL Bleaching on the PLQY Map*

*3.4.1 The Curvature in the Single Pulse Regime*

To consider this, we modified the acquisition scheme in such a way that instead of one long acquisition time per data point, a movie of the same length consisting of several images was recorded. This allows us to monitor the PL intensity evolution during the data acquisition of selected datapoints of the PLQY map.

**Figure** 3a shows a part of the PLQY map with the peculiar "dip" in the single pulse regime for the $MA_{1.1}PbI_{3.1}$ sample (green circles). The same figure also shows the data for the reference MAPI that is horizontal as is to be expected (black diamonds). The other panels of the figure show the time-resolved PL kinetics during the single acquisition of 6 selected data points where the "dip" is observed (*f* varies from 400 Hz to 50 kHz). The green coloration shows the integrated signal that is actually measured and plotted at the PLQY map. The pink coloration shows the expected signal in the case PL did not decrease during the acquisition time. These expected PLQY values are illustrated by the pink line. This data demonstrates that it is the PL dynamics just after switching on the excitation laser that is responsible for the data points being lower than expected at the certain regions of the PLQY map. The behavior of the "corrected" PLQY (pink line) is close to the constant PLQY as expected from the single pulse regime.

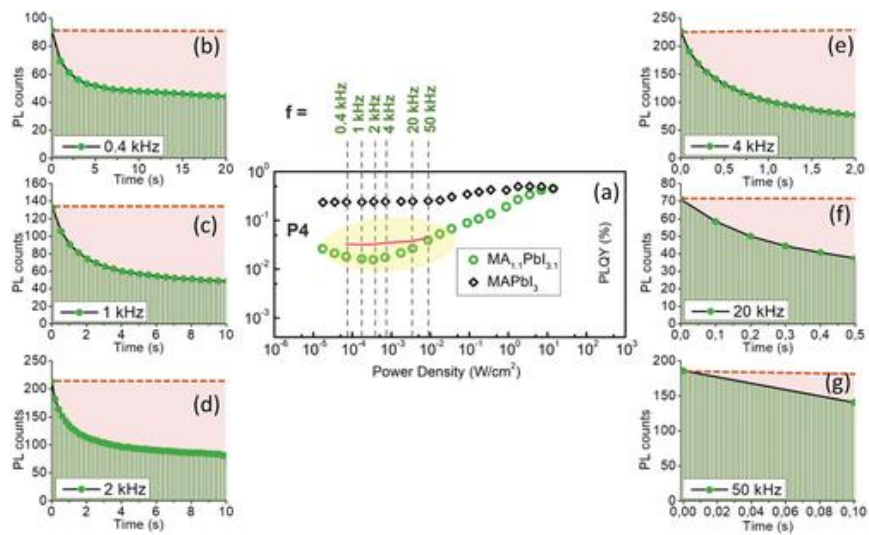

**Figure 3.** a) Dependence of PLQY on repetition rate *f* for the pulse fluence P4 for the significantly overstoichiomeric $MA_{1.1}PbI_{3.1}$ (green circles) and for the stoichiometric $MAPbI_3$ (black diamonds). For frequencies <50 kHz we expect to have single pulse excitation regime implying constant PLQY as the $MAPbI_3$ sample shows. The dependence for the overstoichiomeric sample, however, shows unusual bending down with the lowest value at 2 kHz. This unusual region is highlighted by yellow, see also Figure **1d**. The reason for that is PL kinetics at the time scale of the signal acquisition illustrated in (b–g). The time range for each plot is equal to the signal acquisition time for the corresponding datapoints shown in (a). PL signal evolution is shown in green. The integrals under these curves (green coloration) are proportional to the measured PLQY ((a), green circles). The pink rectangular areas show the "true" PL signal as it would be if the sample was stable. The red line in (a) shows these "true" values which, as expected, form a horizontal line in the given repetition rate range. So, the anomaly in PLQY shown in (a) is due to the PL bleaching at the time scale of seconds shown in (b–g).

Because the acquisition time and power density are different for different data points, and the PL bleaching kinetic does not scale in the same way as the sample exposure time needed for PL collection, the influence of the PL bleaching is different for different regions of the PLQY map. It not only results in the downward shift in the PLQY map (PL bleaching means lower PLQY), but also impacts on its shape that can be immediately detected by visually analyzing the maps. Obviously, the experiment is heavily influenced by the observer effect, i.e., when the excitation light required to induce PL also leads to a change in the material, thus resulting in the change of the observable parameter (PL in our case).

*3.4.2 Jump-Like Artefacts, Effect of Synchronization between the Shutter and the Camera*

Let us consider the region of the map where the average excitation power is quite large and PLQY is also large. In this region the acquisition time needed for obtaining a good signal is very short. Due to the limitation of the shutter installed at the laser beam, the minimum irradiation time of the sample is 70 ms. As one can see in Figure 3, the PL bleaching at high excitation power is very fast with a characteristic time of the order of 100 ms. Importantly, there are two time-intervals that need to be synchronized throughout the experiment: the opening and closing of the shutter blocking the laser (light irradiation interval) and the starting and stopping of the data collection by the CCD camera (data acquisition interval). For a stable sample, the PL signal is just proportional to the overlapping time between these two intervals. However, when the sample is not stable during the measurement, particular details of the timing between the shutter and CCD start to become crucially important (**Figure** 4). Note that these considerations are not limited to PL intensity measurements, they are valid for any optical experiments and careful experimentalists must consider them.

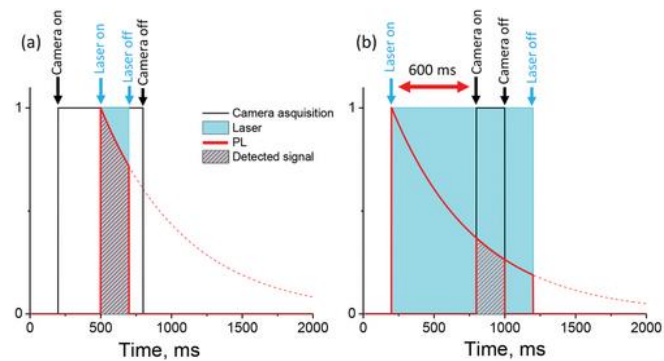

**Figure 4.** A schematic representation of the critical importance of synchronization between the laser irradiation interval and the signal acquisition window. The evolution of PL intensity after opening the laser shutter is shown by the red line. a) Data acquisition in the "fast shutter mode", where the effective data acquisition time is equal to the laser irradiation time as both are determined by the laser shutter. b) Data acquisition in the "slow shutter mode", where there is a delay between the start of the laser illumination and the start of PL acquisition. This delay is 600 ms in our experiments. The acquisition time is 100 ms in both cases, however, the signal detected in (b) is several times smaller than in (a) due to the PL bleaching over the first 600 ms after opening the laser shutter.

The jump-like artefacts observed for non-stoichiometric samples at several positions on their PLQY maps originate from the issue mentioned above. The reason for the drastic jumps of the PLQY from one data point to another observed in some regions of the PLQY map is that they originate from the two different regimes of the data acquisition in terms of the synchronization between the laser irradiation and PL acquisition time intervals.

Figure 4 illustrates these two different modes used in the measurements. The reason for having these two different modes in the setup was purely historical, as it is common in any setup/measurement routine that is constantly under development in a research lab. The two modes being: a) the "fast shutter" mode, where the sample irradiation and PL collection intervals are identical (perfect synchronization, was used for irradiations times <100 ms), and b) the "slow shutter" mode where the sample irradiation time is longer than the PL acquisition time (delayed synchronization, was used when the irradiation time >100 ms). It is obvious that if PL is bleaching at the time scale comparable with the synchronization delay (600 ms), the PL intensity measured with the "slow shutter" should be smaller than that measured with the "fast shutter" for the same PL acquisition time, as represented in Figure 4c. However, when the acquisition time is >1 s, there should be no difference between these two modes.

The easily observed reproducible artefacts visible in the PLQY maps are therefore due to sensitivity of the overstoichiometric samples to light. In fact, we do not see any real possibility to eliminate these artefacts. The observer effect is always there. We can only conclude that by applying automatic measurement routines, for example, PLQY($f$,P) mapping, it is possible to identify and classify samples by their photosensitivity.

*3.5 Time Evolution of the PLQY Maps during Long-Term Storage*

Having identified the reasons for the artefacts in the PLQY maps, we examine how these maps change upon storage of the samples. We always used the same data acquisition routine, that is why the maps of different samples and the same samples measured after different storage time can be directly compared. We find that the artefacts due to the defect dynamics in overstoichiometric samples decrease with the age of the films (**Figure** 5). The most drastic changes are for $MA_{1.04}PbI_{3.04}$. As the PLQY maps show, PLQY increases with age and the jumps due to fast dynamics at high excitation power also decrease. Although the PLQY map for the 6 weeks old sample still shows artefacts like non-constant PLQY for the single pulse regime, it has become much closer to the reference stoichiometric MAPI (compare with Figure 1b). This dramatic change in the PL response signifies healing of the sample. We hypothesize that the excessive iodide in the crystal structure (I⁻ interstitials inherent to overstoichiometric samples), first diffuse to the surface of the grains and then leave the surface over time upon exposure to oxygen and light illumination, see also the discussion in ref. [16] where healing of an overstoichiometric MAPI film was observed after deliberate photo-oxidation.

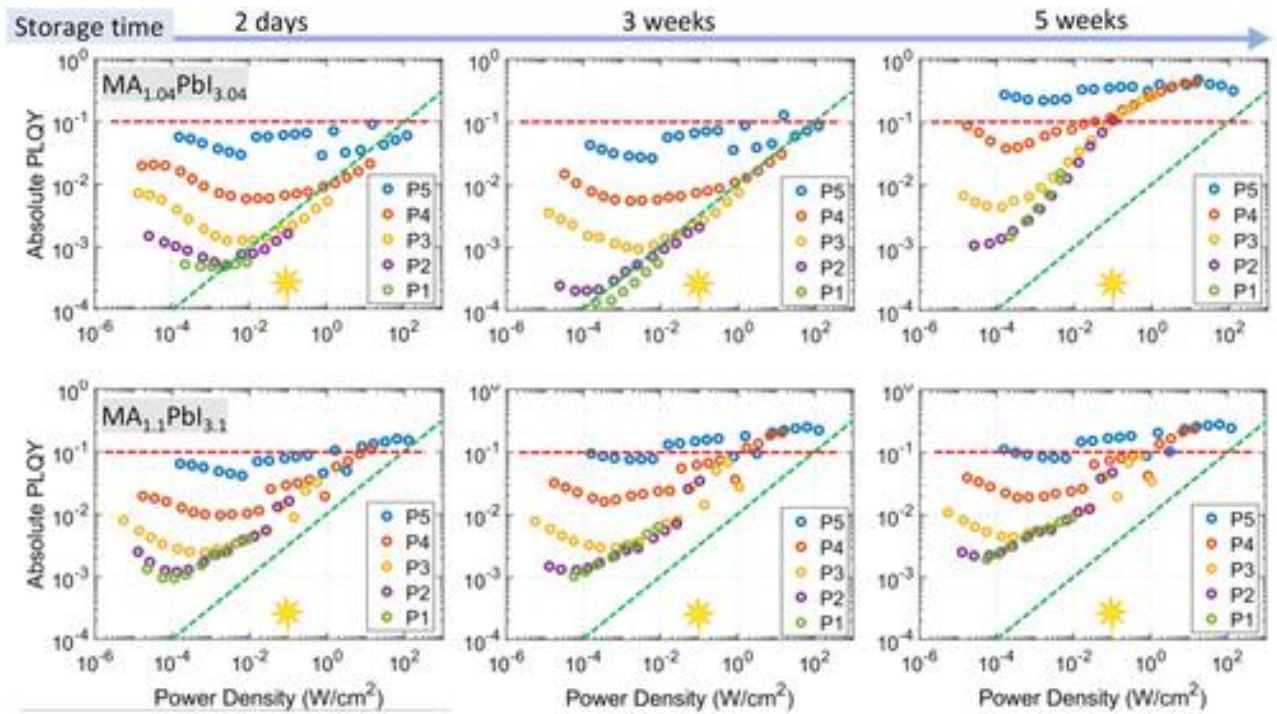

**Figure 5.** A series of PLQY($f$,P) maps demonstrating the evolution of the response of the non-stoichiometric MAPI films as a function of the storage time after synthesis. Top raw—the slightly over-stoichiometric $MA_{1.04}PbI_{3.04}$, bottom raw—the significantly over-stoichiometric $MA_{1.1}PbI_{3.1}$ films. Red horizontal dash line shows PLQY = 10%, the tilted green dash line is the square root dependence (slope 0.5 in log-log scale) expected for the quasi-CW regime for the case of photodoping without trap saturation. These lines aid comparison of the plots between each other. The "sun" marks the excitation power density of 0.1 W cm$^{-2}$.

Evolution of the PLQY maps for the significantly over-stoichiometric $MA_{1.1}PbI_{3.1}$, is shown in the bottom row, in Figure 5d–f. Although the PLQY value also slightly goes up as the sample ages, the artefacts due to fast light-induced defect dynamics remain the same. It means that when the degree of overstoichiometry is over a certain limit, the sample cannot spontaneously transform to its normal state. Even after long storage time it retains the charge carrier and defect dynamics significantly different from that of the stoichiometric MAPI film.

**3.6 Possible Mechanisms and Origin of Defect Photosensitivity and Sample Maturing**

To date there is no complete understanding about which particular defects are responsible for the apparent photosensitivity of MAPI exemplified in Figure 2. Note that we were not able to find any publication where PL evolution of MHPs on the sub-second time scale has been demonstrated. Our results as well as the literature data[18, 25, 26, 36] indicate that iodide interstitials inherently present at high concentrations in overstoichiometric samples is at least one of the factors. The corresponding excess of $MA^+$ counter ions also exist in the overstoichiometric samples; however, they are believed to be much less important for non-radiative recombination processes.[37]

Here we would like to make a connection to the mixed halide perovskites (e.g., Br/I mixtures[38, 39]) where the steady state of the material in terms of its bandgap and many other properties depends on the light (intensity, repetition rate for pulsed excitation[40, 41]) the sample is irradiated by. Although it is easy to study phase segregation in these systems experimentally, even for the model $MAPbI_xBr_{3-x}$ perovskite the full physical picture of the processes is not yet established.[38] In general, steady state of any photosensitive material with reversible photo-induced reactions (mixed halides MHPs as a very clear case and, as we propose here, MHPs in general) is determined by kinetic equilibrium between several processes involving various reactants/products and photons (or electronic excitations). Changing of light irradiation condition inevitably changes the resulting equilibrated state. We speculate that in MAPI films initially containing lots of defects ($I^-$ and $MA^+$ interstitials) irradiation by light intensifies the defect migration. This migration leads to spatial redistribution of ions and may lead to local segregation of the material in terms of defect concentration/local composition that results in enhanced non-radiative charge recombination in overstoichiometric samples. This material state exists in a dynamic equilibrium between light-driven segregation and gradient driven diffusion in analogy to mixed halide perovskites. In this model, the degree of segregation depends on the light intensity and so does the equilibrated value of PL.

**3.7 The Observer Effect as a Problem for Understanding Perovskites**

The reversible fast PL bleaching process discussed in this work can be easily overlooked because it happens on time scale of seconds while in most of experiments steady state values of parameters (in terms of the data acquisition time) are measured. Indeed, commonly the sample is considered unstable and not suitable for measurements only if the experimentally measured parameter in the second experiment is different (e.g., PL intensity is lower) from that measured in the first and so on. However, this simple test is not enough for photosensitive perovskites. Due to self-healing the measured parameters in the second experiment most probably will not be different from the first one. It means that the sample appears as "stable" as soon as the same experimental routine is applied again and again.

Indeed, the temporal bleaching we discuss here is not related to permanent degradation, but to the equilibrium state of the material under light irradiation. We observe that PL intensity changes when the material transfers to a new equilibrium when the irradiation conditions are changed (e.g., the laser was switched on, or the intensity of the laser is changed). As soon as the new equilibrium is reached, the PL (and therefore, the defect concentration) stabilizes. Whether or not the experimentalist has a chance to notice this instability depends on the interplay of multiple experimental conditions such as the sample relaxation time, data acquisition time, gaps between measurements, type of irradiation (constant, interrupted) and many other experimental details.

The problem with these steady-state data is that the quasi-stable equilibrium condition of the sample depends on the light intensity, which means that obtained steady-state parameters cannot be compared with the same parameters measured at different light irradiation conditions. In other words, changing the irradiation conditions is equivalent to changing the sample. This problem undermines the commonly applied method of varying photogenerated charge carrier concentration as the universal approach for understanding charge carrier dynamics.

Thanks to the automated experimental routine and the unique mapping in very large $f$ and P space, we can immediately identify that some samples exhibit unusual behavior and should be not even tried to be analyzed without serious consideration of their photosensitivity. Strictly speaking, for materials with photosensitive defect concentration standard types of spectroscopic methods and theoretical models might not be fully applicable.

## 4 Conclusion

Automatically recorded PLQY($f$,P) maps serve as easily readable fingerprints of sample electronic properties. Such a fingerprint depends on charge carrier recombination pathways which to a large extent are controlled by defect state properties and their concentration that can be strongly photosensitive. Photosensitivity of defect concentration is very pronounced in overstoichiometric films having excess of $MA^+$ and $I^-$ in their composition. This was revealed in the PLQY($f$,P) maps as clear artifacts beyond any sensible explanation in the framework of standard charge recombination models. These films show a strong observer effect, meaning that by performing an optical experiment one changes the sample and thus the results reflect properties of the material convoluted with the details of the experimental routine used in the particular laboratory. The reason behind these artifacts is that PL intensity reversibly declines under light exposure with characteristics timescale from hundreds of milliseconds to tens of seconds depending on light excitation conditions. Following these artifacts over the storage time of the film over weeks, we find that slightly overstoichiometric films change their properties toward being more stable, having a higher PLQY and PLQY($f$,P) maps closer to the one for the standard MAPI. This phenomenon is one more example of the self-healing ability of metal halide perovskite semiconductors enabled by efficient ion diffusion in the crystal. The observer effect and self-healing must be considered in any optical and electronic measurement of metal halide perovskites.

**Acknowledgements**

The work was supported by the Swedish Research Council (2020-03530). Q.A. and Y.V. thank the center for advancing electronics Dresden (cfaed) for generous funding in the framework of the postdoc program and the Deutsche Forschungsgemeinschaft (DFG) for funding the "PERFECT PVs" project (Grant No. 424216076) via the Special Priority Program 2196. Y.V. has received funding from the European Research Council (ERC) under the European Union's Horizon 2020 research and innovation programme (ERC grant agreement number 714067, ENERGYMAPS).

Perspective. *Energy Environ. Sci.* **2020**, 2024–2046.